\documentclass[twocolumn,showpacs,preprintnumbers,amsmath,amssymb]{revtex4}

\usepackage{graphicx}
\usepackage{dcolumn}
\usepackage{bm}

\begin{document}

\preprint{APS/123-QED}

\title{Approximation scheme for master equations: variational approach to multivariate case}

\author{Jun Ohkubo}
\email[Email address: ]{ohkubo@issp.u-tokyo.ac.jp}
\affiliation{
Institute for Solid State Physics, University of Tokyo, 
Kashiwanoha 5-1-5, Kashiwa-shi, Chiba 277-8581, Japan
}
\date{\today}

\begin{abstract}
We study an approximation scheme based on a second quantization method for a chemical master equation.
Small systems, such as cells, could not be studied by the traditional rate equation approach
because fluctuation effects are very large in such small systems.
Although a Fokker-Planck equation obtained by the system size expansion includes the fluctuation effects,
it needs large computational costs for complicated chemical reaction systems.
In addition, discrete characteristics of the original master equation are neglected in the system size expansion scheme.
It has been shown that the usage of the second quantization description and a variational method
achieves tremendous reduction in the dimensionality of the master equation approximately,
without loss of the discrete characteristics.
We here propose a new scheme for the choice of variational functions,
which is applicable to multivariate cases.
It is revealed that the new scheme gives better numerical results than old ones
and the computational cost increases only slightly.
\end{abstract}

\maketitle

\section{Introduction}

Time evolution of chemical reaction systems can be studied by a traditional rate equation approach.
In the rate equation approach, all fluctuation effects are neglected.
In order to treat the chemical reaction systems more appropriately,
it is possible to write discrete master equations which include all information of the fluctuation effects.
In recent years, it has been necessary to study a stochastic process in a system with \textit{small size}.
Here, the word `small size' means that the number of molecules contained in the system is very small.
In such small size system, the fluctuation effects are large and cannot be ignored.
For example, some of biochemical systems, such as gene regulatory networks, 
contain small number of regulatory molecules \cite{Elowitz2002,Sasai2003}.
It is expected that the fluctuation effects and the discrete effects of the original master equation
play an important role in such small size systems.

In general, we cannot solve the chemical master equations analytically.
While one might think that direct numerical evaluation of the chemical master equation
is the best way in order to investigate such small systems, it would be also impossible.
The master equation can be understood as a large set of coupled ordinary differential equations,
and the number of them often becomes very large;
it increases exponentially with the number of chemical substances.
For example, we have $100^5 = 10^{10}$ coupled differential equations 
when we consider a small size system of only five chemical substances with the number of each chemical substances
varying between $0$ and $99$.
It is possible to approximate the chemical master equations by a Fokker-Planck equation
which is obtained by the system size expansion method \cite{Risken1984,Gardiner2004}.
However, the Fokker-Planck equation has the similar problem of the \textit{curse of the dimensionality},
i.e., the computational costs increases exponentially with the number of chemical substances.
In addition, the system size expansion method neglects the discrete characteristics of the original chemical master equation,
and the system size expansion method becomes a good approximation only if we consider a chemical system with large size.

A chemical reaction system consists of Markov jump processes in continuous time.
Such dynamics can be simulated (exactly) using several Monte Calro methods such as the Gillespie algorithm
\cite{Gillespie1977}.
In the Gillespie algorithm, the lapse time to the next event is determined 
by exponentially distributed random numbers,
and we determine which event occurs in proportion to the rate of the event.
While the numerical simulations are available for studying complicated stochastic systems,
these methods could sometimes become time-consuming when there are many chemical reactions,
because averaging procedures of a lot of Monte Carlo samples are needed in order to obtain statistics values.
Hence, it is important to study analytical methods and approximation schemes applicable to small systems.

The field theoretic approach (or second quantization description)
is one of the candidates for such analytical treatments.
The analogy of the master equation to quantum descriptions has been introduced by Doi, 
and several authors have developed the formalism \cite{Doi1976,Doi1976a,Peliti1985}.
The field theoretic approach has revealed the anomalous kinetics in reaction-diffusion systems incorporating 
the renormalization group method \cite{Mattis1998,Tauber2005}, 
and the analytical scheme has been applied to various phenomena
\cite{Pigorsch2002,Dickman2002,Dickman2002a,Stilck2004,Buice2007}.
In addition, the variational method based on the second quantization description 
developed by Sasai and Wolynes \cite{Sasai2003}
is hopeful in order to investigate the fluctuation and discrete effects in the small size systems.
The variational method is an approximation, and then we can avoid the curse of dimensionality.
In addition, the method does not ignore the discrete characteristics of the original chemical master equations.

The variational method has been already applied to stochastic gene regulatory networks,
and it has given qualitatively good results \cite{Sasai2003,Kim2007,Kim2007a}.
However, the original formulation of the variational method has been based on the Poisson ansatz,
in which we assume that the probability distribution of the system is described by the Poisson distribution.
Hence, it is necessary to use a Hartree approximation, 
and then only restricted fluctuation effects can be included \cite{Kim2007,Kim2007a}.
In order to improve the variational method,
``a superposition ansatz'' has been proposed in ref.~\cite{Ohkubo2007}.
The new ansatz is based on the coherent state,
and gives a quantitatively good result for master equations describing the gene regulatory network.

The main purpose of the present paper is to investigate the advantage of the new variational ansatz 
and its application limitation.
In ref.~\cite{Ohkubo2007}, the validity of the superposition ansatz
has been confirmed for a \textit{one-body problem} in which the Hartree approximation is adequate.
However, the applicability of the superposition ansatz for \textit{multivariate cases} has not been studied.
For practical purposes, the multivariate case is important for real chemical reactions in cell.
Hence, in the present paper, we mainly focus on the applicability of the new ansatz to multivariate problems.
We first consider a simple univariate problem in order to investigate zero-boundary effects:
the reaction system stops when the number of molecules becomes zero.
For the univariate problem,
we clarify that the survival probability is easily calculated from the Laplace transformation of the variational function.
Next, we consider a bivariate problem which cannot be treated by the original Poisson ansatz adequately.
For the bivariate problem, we will show that the Hartree approximation does not give quantitatively good results
in a chemical reaction system with small size.
The superposition ansatz gives better results than those of the Hartree approximation.
In addition, it enables us to calculate the covariance,
which is incomputable by the Hartree approximation.

The construction of the present paper is as follows.
In Sec.~II, we will explain the formalism of the second quantization method and the variational method.
Section III gives numerical results for a univariate problem,
and we will show an easy calculation scheme for the survival probability.
In Sec.~IV,
we treat a bivariate problem and study the availability of the superposition ansatz.
Section~V gives concluding remarks.

\section{Formalism}

At first, we summarize the second quantization description for a master equation
and the variational method developed in refs~\cite{Sasai2003} and \cite{Ohkubo2007}.

\subsection{Second quantization description}

We consider a system with $d$ chemical substances, $X_1, \dots, X_d$,
and denote the number of each chemical substance $X_i$ as $n_i$.
Each reaction has the following generic structure:
\begin{align}
\sum_j \nu_j^a X_j \stackrel{c^a}{\to} \sum_j \eta_j^a X_j,
\end{align}
and hence the probability of a transition per unit time is proportional to the rate $c^a$
and the number of molecules, giving a reaction rate of
\begin{align}
R^a(\{n_i\}) = c^a \prod_j \frac{n_j!}{(n_j - \nu_j^a)!},
\end{align}
where $\{n_i\} \equiv \{n_1, n_2, \dots, n_d\}$.
In this case, the master equation for the probability of a configuration $\{ n_i \}$ has the form:
\begin{align}
\frac{d}{dt} P( \{ n_i \} ) = \sum_a [R^a(\{n_i^a\}) P(\{n_i^a\}) - R^a(\{n_i\}) P(\{n_i\}),
\label{eq_master}
\end{align}
where $n_i^a \equiv n_i + \nu_i^a - \eta_i^a$
is the number of molecules prior to reaction $a$ that leads to a current number $n_i$.

In general, it is useful to rewrite the master equation~\eqref{eq_master} 
by creation and annihilation operators and to treat it by techniques developed by quantum mechanics.
First of all, we define a ket vector $| \{n_i\} \rangle$ as the state in which 
the configuration of the system is $\{n_i\}$.
We here introduce the bosonic operator algebra
\begin{align}
[ a_i, a_j^\dagger ] = \delta_{ij}, [a_i, a_j] = [a_i^\dagger, a_j^\dagger] = 0,
\end{align}
where $a_i^\dagger$ is the creation operator for chemical substance $X_i$,
and $a_i$ the annihilation operator for $X_i$.
The creation and annihilation operators act on the ket vector $| \{n_i\} \rangle = | n_1, \dots, n_d \rangle$
as follows:
\begin{align}
&a_i^\dagger | n_1, \dots, n_i, \dots, n_d \rangle = | n_1, \dots, n_i + 1, \dots, n_d \rangle, \\
&a_i | n_1, \dots, n_i, \dots, n_d \rangle = n_i | n_1, \dots, n_i - 1, \dots, n_d \rangle, \\
&a_i^\dagger a_i | n_1, \dots, n_i, \dots, n_d \rangle = n_i | n_1, \dots, n_i, \dots, n_d \rangle. 
\end{align}
Therefore a state with the configuration $\{ n_1, \dots, n_d \}$
is obtained from the empty vacuum state $| 0 \rangle$ as
\begin{align}
| \{ n_i \} \rangle = \prod_{i=1}^d (a_i^\dagger)^{n_i} | 0 \rangle,
\end{align}
where the empty vacuum state $| 0 \rangle$ is defined as $a_i | 0 \rangle = 0$ for all $i$.

Using the time-dependent probability distribution of the original discrete master equation, $P(\{n_i\})$,
we introduce the ket vector of the system as
\begin{align}
| \Psi \rangle = \sum_{\{n_i\}} P(\{ n_i \}) | \{n_i \} \rangle.
\end{align}
The original master equation is then translated into the following equation
\begin{align}
\frac{\partial}{\partial t} | \Psi \rangle = \Omega | \Psi \rangle,
\label{eq_master_equation_rev}
\end{align}
where $\Omega$ is the time evolution operator.
The time evolution operator $\Omega$ is constructed so that it becomes consistent with the original master equation.
We will show concrete examples of the time evolution operator $\Omega$ in Secs.~III and IV.

We remark on the computational method for the calculation of averages
with respect to the original probability distribution $P(\{n_i\})$.
The average is taken by the usage of the projection state $\langle \mathcal{P}| \equiv \langle 0 | \prod_{i=1}^d \exp(a_i)$.
The projection state satisfies $\langle \mathcal{P} | \Psi \rangle = 1$ and 
a statistical average of an observable $A$ is obtained by
\begin{align}
\langle A \rangle = \sum_{\{n_i\}} A(\{n_i\}) P(\{n_i\})
= \langle \mathcal{P} | A(\{a_i^\dagger a_i\}) | \Psi \rangle.
\end{align}

Using the above theoretical scheme, 
we can calculate various quantities for the original chemical master equation \textit{rigorously}.
However, the evaluation of the time dependent ket vector $| \Psi \rangle$ is a computationally difficult task,
and in general it is impossible to obtain the exact solution of the ket vector.
We, therefore, use an approximation method in order to obtain the time-evolution of the ket vector.

\subsection{Variational method}

In order to reduce the dimensionality of the problem,
a variational method developed by Eyink \cite{Eyink1996,Alexander1996} is available.
We here briefly review the variational method combined with the second quantization description \cite{Sasai2003}.

When we define an effective action $\Gamma$ as
\begin{align}
\Gamma = \int dt \langle \Phi | (\partial_t - \Omega) | \Psi \rangle,
\end{align}
Eq.~\eqref{eq_master_equation_rev} is equivalent to the functional variation
$\delta \Gamma / \delta \Phi = 0$.
Because of the non-Hermitian property,
it is not always true that the left eigenvectors and right eigenvectors are the same.
Hence, we assume two variational functions for the bra and ket states, respectively:
The ket state $| \Psi \rangle$ (or the bra state $\langle \Phi |$) 
is parametrized by $\alpha^\mathrm{R}$ (or $\alpha^\mathrm{L}$),
and where $\alpha^\mathrm{R}$ and $\alpha^\mathrm{L}$ are vectors with $K$ components;
\begin{align}
 \alpha^\mathrm{R} &= \{ \alpha_1^\mathrm{R}, \alpha_2^\mathrm{R}, \cdots, \alpha_K^\mathrm{R}\}, \\
 \alpha^\mathrm{L} &= \{ \alpha_1^\mathrm{L}, \alpha_2^\mathrm{L}, \cdots, \alpha_K^\mathrm{L}\}.
\end{align}
A set of finite dimensional equations for parameters $\alpha^\mathrm{R}$ and $\alpha^\mathrm{L}$ is obtained 
by the functional variation procedure.
Note that we set $\Phi(\alpha^\mathrm{L} = 0 )$ to be consistent with the probabilistic interpretation,
so that
\begin{align}
\langle \Phi(\alpha^\mathrm{L} = 0 ) | \Psi ( \alpha^\mathrm{R} ) \rangle = 1.
\end{align}
We, therefore, obtain the following equation which stems from an extremum of the action
\begin{align}
\left[
\sum_{l=1}^K \left\langle \frac{\partial \Phi}{\partial \alpha_{m}^\mathrm{L}} \right|
 \left. \frac{\partial \Psi}{\partial \alpha_{l}^\mathrm{R}} \right\rangle
\frac{d \alpha_{l}^\mathrm{R}}{d t}
- \left\langle \left. \left. \frac{\partial \Phi}{\partial \alpha_{m}^\mathrm{L}} \right| \Omega
\right| \Psi \right\rangle 
\right]_{\alpha_{m}^\mathrm{L} = 0} = 0 \notag \\
 \quad \mathrm{for} \,\, m = 1,2,\cdots, K.
\label{eq_variational_equation}
\end{align}
Using this variational scheme, we obtain a set of time-evolution equations for 
the time-dependent parameters $\alpha^\mathrm{R}$,
and the equations can be solved numerically.
The only remaining task is to give an explicit ansatz for $\langle \Phi |$ and $| \Psi \rangle$.

\subsection{Poisson ansatz}

The Poisson ansatz has been introduced in ref.~\cite{Sasai2003}.
In the Poisson ansatz, we assume that the probability of the number of each molecule, $n_i$,
obeys the Poisson distribution.
To be more precise,
the probability distribution of the original discrete master equation assumes to have a product form
\begin{align}
&P(\{n_i\})_\mathrm{Poisson} = \frac{e^{-\mu_1} \mu_1^{n_1}}{n_1 !} \dots \frac{e^{-\mu_d} \mu_d^{n_d}}{n_d !},
\end{align}
where $\mu_i$ is the mean value of the number of molecule $X_i$.
This means that we need a Hartree approximation 
\begin{align}
| \Psi \rangle = | \psi_1 \rangle \otimes \dots \otimes | \psi_d \rangle,
\end{align}
and each one-body ket vector $| \psi_i \rangle$ is assumed as
\begin{align}
| \psi_i \rangle = \exp\left[ \mu_i (a_i^\dagger-1) \right] | 0_i \rangle,
\end{align}
where $| 0_i \rangle$ is the vacuum state for a chemical substance $X_i$.
In this case, we have $d$ time-dependent parameters $\alpha^\mathrm{R} = \{ \mu_1, \dots, \mu_d \}$,
and by using an adequate bra ansatz,
$d$ coupled differential equations for the time-dependent parameters $\{\mu_i\}$ are obtained.

We note that the Poisson ansatz includes only restricted fluctuation effects
because the mean value and the variance of the Poisson distribution are the same.
In addition, we cannot calculate any correlations between chemical substances 
because the Poisson ansatz needs the Hartree approximation.

\subsection{Superposition ansatz}

We here use the analogy between the Poisson ansatz and the coherent state of quantum mechanics.
A new ansatz, ``superposition ansatz,'' is constructed by the superposition of the coherent states as follows:
\begin{align}
| \Psi \rangle = 
\int_0^\infty d \bm{x} \,  h(\{ x_i \}; \{ \bm{\mu} \} ) \prod_{j=1}^d \exp[x_j (a_j^\dagger -1)] | 0 \rangle,
\label{eq_superposition}
\end{align}
where $\int_0^\infty d \bm{x} \equiv \int_0^\infty dx_1 \dots \int_0^\infty dx_d$,
and $h(\{ x_i \}; \{ \bm{\mu} \} )$ is the variational function.
Here, $\{\bm{\mu}\}$ is a set of the variational parameters 
which specifies the variational function $h(\{ x_i \}; \{ \bm{\mu} \} )$.
Note that it is possible to use a \textit{continuous} probability distribution as the variational function $h$
without loss of the discrete characteristics of the original problem, via the usage of the coherent states.
In general, the continuous property sometimes makes analytical treatments easier than discrete cases.
In addition, the superposition ansatz enables us to use a multivariate variational function $h(\{ x_i \}; \{ \bm{\mu} \} )$
so that the correlations between the chemical substances can be adequately calculated by the new ansatz.

\section{Example I: univariate case}

First, we study a univariate case.
In this simple example, we mainly focus on the zero-boundary effect:
the number of molecules cannot be below zero.
When we use the Fokker-Planck equation,
the additional boundary condition is needed in order to include the zero-boundary effect.
In contrast, the variational method does not need to consider the zero-boundary,
because the superposition ansatz is based on the Poisson distributions (coherent states)
and hence the discrete characteristics is not neglected.
In addition, we will show that the survival probability is easily calculated
by the Laplace transformation of the variational function in the superposition ansatz.

\subsection{Model}

We consider the following reaction system:
\begin{align}
\begin{array}{l}
X \xrightarrow{c_1} 2X,  \\
X \xrightarrow{c_2} \varnothing. \\
\end{array}
\end{align}

The master equation is given by
\begin{align}
\frac{d}{dt} P_{n_x} =& c_1 \left[ (n_x - 1) P_{n_x-1}  - n_x P_{n_x} \right] \notag \\
&+ c_2 \left[ (n_x+1) P_{n_x+1} -  n_x P_{n_x} \right] ,
\label{eq_uni_master}
\end{align}
where $n_x$ is the number of molecules $X$.
Using the second quantization description,
we finally obtain the following time evolution operator:
\begin{align}
\Omega = c_1 (a_x^\dagger a_x^\dagger a_x - a_x^\dagger a_x)
+ c_2 (a_x - a_x^\dagger a_x).
\end{align}
It is easy to check the above time evolution operator adequately recovers the original master equation
\eqref{eq_uni_master}.

We here use the gamma distribution as the continuous variational function 
$h(\{x_i\},\{\bm{\mu}\})$ in Eq.~\eqref{eq_superposition}.
The gamma distribution is defined as
\begin{align}
F(x;k,\theta) = x^{k-1} \frac{\exp(-x/\theta)}{\Gamma(k) \theta^k},
\end{align}
and the mean and the variance are given by $k\theta$ and $k \theta^2$, respectively.
This choice of the variational function indicates 
that the variational parameters $\{\bm{\mu}\}$ are given by $\{\mu_1, \mu_2\} = \{k, \theta\}$.
The first and second moments of the gamma distribution are calculated as
\begin{align}
\langle x \rangle &= k\theta, \label{eq_uni_first}\\
\langle x^2 \rangle &= k\theta^2 + k^2 \theta^2, \label{eq_uni_second}
\end{align}
respectively.
Note that the above first and second moments \textit{does not} correspond to those of $n_x$.
The gamma distribution is introduced for the variational function,
and hence the real probability distribution is obtained by the superposition 
of the Poisson distributions weighted by the gamma distribution.
The explicit forms of the mean and variance of $n_x$ will be given in the next subsection.

By using the following ket ansatz and bra ansatz
\begin{align}
| \Psi \rangle = \int_0^\infty dx F(x; k, \theta)
\exp[ x (a_x^\dagger -1) ] | 0_x \rangle,  \\
\langle \Phi | = \langle 0_x | \exp\left\{ a_x + \lambda^{(1)} a_x + \lambda^{(2)} (a_x)^2 \right\},
\end{align}
the variational parameters are 
\begin{align}
\begin{array}{l}
\alpha^\mathrm{R} = \{ k, \theta \}, \\
\alpha^\mathrm{L} = \{ \lambda^{(1)}, \lambda^{(2)} \}. \\
\end{array}
\end{align}
From Eq.~\eqref{eq_variational_equation},
we finally obtain the following coupled differential equations:
\begin{align}
&\frac{d k}{dt} \frac{\partial}{\partial k} \langle x \rangle
+ \frac{d \theta}{dt} \frac{\partial}{\partial \theta} \langle x \rangle
= c_1 \langle x \rangle - c_2 \langle x \rangle 
\label{eq_uni_evolution_1},\\
&\frac{d k}{dt} \frac{\partial}{\partial k} \langle x^2 \rangle
+ \frac{d \theta}{dt} \frac{\partial}{\partial \theta} \langle x^2 \rangle 
= c_1 \left( 2 \langle x^2 \rangle + 2 \langle x \rangle \right) 
- 2 c_2 \langle x^2 \rangle
\label{eq_uni_evolution_2}.
\end{align}
Note that the first and second moments, $\langle x \rangle$ and $\langle x^2 \rangle$
depends on the variational parameters $k$ and $\theta$, as described in Eqs.~\eqref{eq_uni_first}
and \eqref{eq_uni_second},
and then the derivatives, e.g., $\partial \langle x^2 \rangle / \partial k$,
are calculated explicitly.

\subsection{Numerical result}

We compare numerical results obtained by the superposition ansatz with those of the Gillespie algorithm.
For simplicity, we here set the reaction rates as $c_1 = c_2 = 1$.
The initial parameters are set as $k_\textrm{ini} = 30.0$ and $\theta_\textrm{ini} = 0.1$.
When we use the variational method, the only necessary thing is to evaluate 
the time evolution of the variational parameters $k$ and $\theta$ by using Eqs.~\eqref{eq_uni_evolution_1}
and \eqref{eq_uni_evolution_2}.
However, in the Gillespie algorithm, the averaging procedure for the Monte Carlo results is needed
and we took the averages over $10^5$ runs.

Because we set $c_1 = c_2$, the mean $\langle n_x \rangle$ does not change with time.
On the other hands, the variance increases in proportional to time.
From the variational method, we obtain the time evolution of the variational parameters $k$ and $\theta$ numerically
(from Eqs.~\eqref{eq_uni_evolution_1}, \eqref{eq_uni_evolution_2}).
Using these variational parameters,
the variance of $n_x$ is calculated as
\begin{align}
\textrm{Var}[n_x] = \langle n_x^2 \rangle - \langle n_x \rangle^2 = \langle x \rangle + \langle x^2 \rangle - \langle x \rangle^2
= k\theta + k \theta^2,
\label{eq_variance}
\end{align}
because
\begin{align}
\langle n_x^2 \rangle &= \sum_{n_x} n_x^2 \int_0^\infty dx F(x;k,\theta) \frac{e^{-x} x^{n_x}}{n_x!} \notag \\
&= \int_0^\infty dx x(1+x) F(x;k,\theta) \notag \\
&= \langle x \rangle + \langle x^2 \rangle.
\end{align}
In the similar way, it is easy to show the following relationship:
\begin{align}
\langle n_x \rangle &= \langle x \rangle.
\end{align}

\begin{figure}
\begin{center}
  \includegraphics[width=70mm,keepaspectratio,clip]{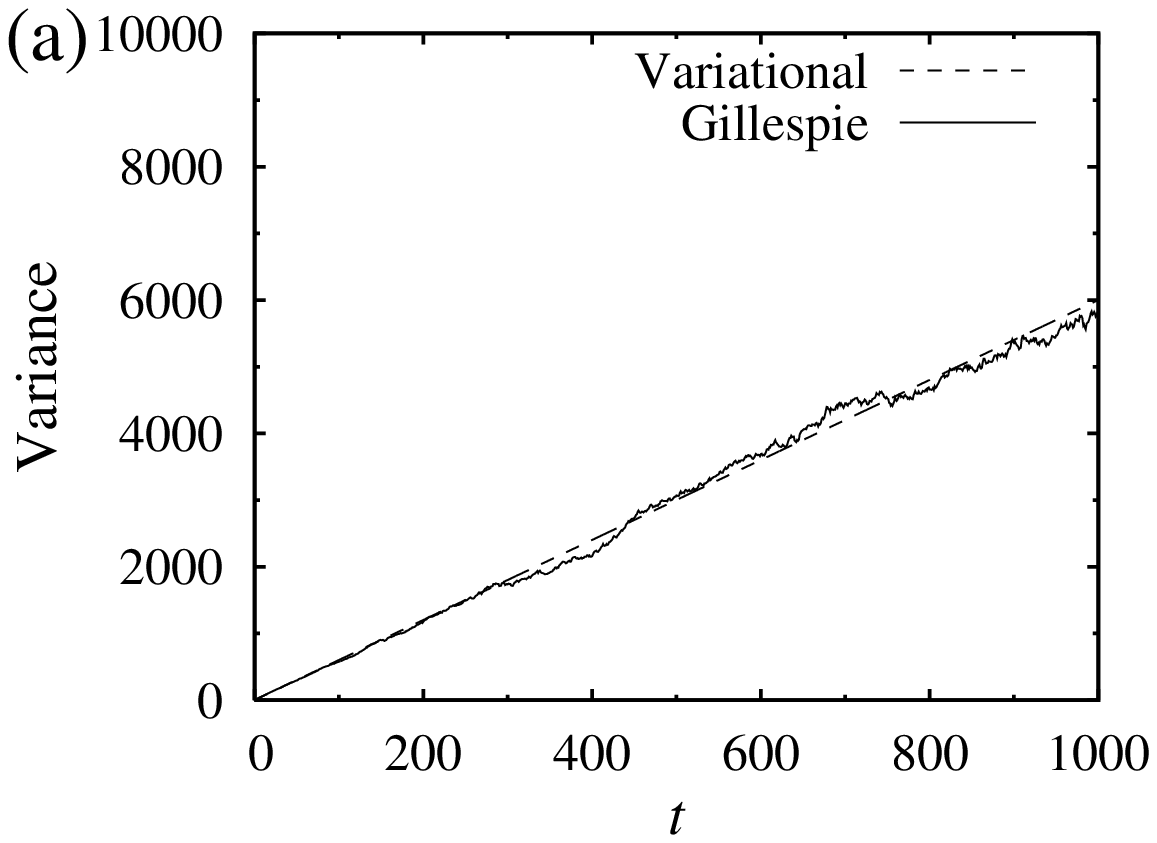} \\
  \includegraphics[width=70mm,keepaspectratio,clip]{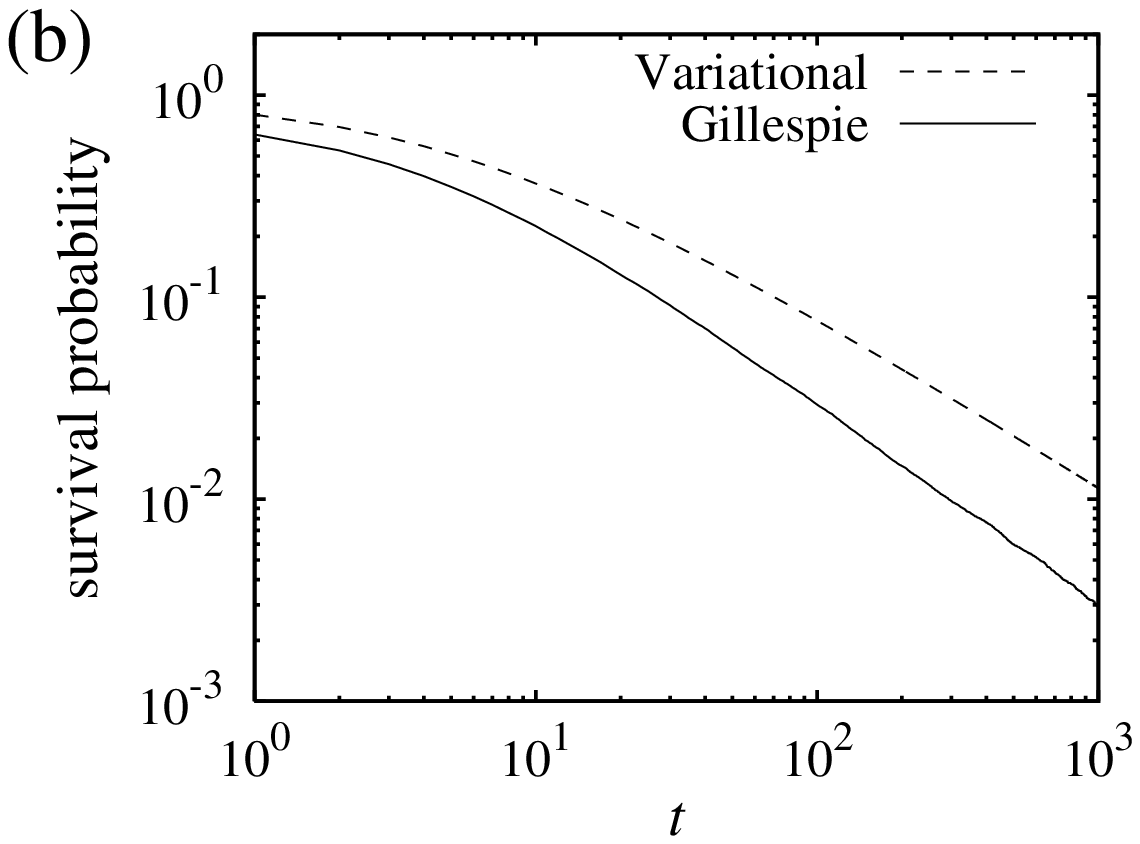} 
\caption{
(a) The time evolution of the variance (Eq.~\eqref{eq_variance}).
(b) The survival probability (Eq.~\eqref{eq_survival}).
}
\label{fig_result_uni}
\end{center}
\end{figure}

Figure~\ref{fig_result_uni} shows the numerical results
obtained by the variational method and the Gillespie algorithm.
The variance of $n_x$ is shown in Fig.~\ref{fig_result_uni}(a).
One can see that the results obtained by the variational method and the Gillespie algorithm
are in good agreement.
Note that the time evolution of the variance cannot be calculated by the Poisson ansatz,
because the variance of the Poisson ansatz is restricted to the same value as the mean.
Next, we calculate the survival probability.
The survival probability is defined as the probability with which the number of molecules is not zero.
In the variational method, it is easy to calculate the survival probability:
In a Poisson distribution with mean $x$,
the probability with which the number of molecules is zero is given by
\begin{align}
\left. e^{-x} \frac{x^k}{k!} \right|_{k=0} = e^{-x}.
\end{align}
In the superposition ansatz, the whole probability distribution is given by
the superposition of the Poisson distribution with mean $x$ weighted by the
variational function.
Hence, by using the Laplace transformation of the variational function $F(x;k,\theta)$,
we obtain the survival probability as follows:
\begin{align}
P_\textrm{surv} \equiv \int_0^\infty dx F(x; k,\theta) e^{-x} = \left(1+\frac{1}{\theta} \right)^{-k} \theta^{-k}.
\label{eq_survival}
\end{align}
In Fig.~\ref{fig_result_uni}(b), we show the numerical results of the survival probability.
We get a qualitatively good result from the variational method;
the survival probability shows slow decay.
In the variational scheme, we construct the probability distribution based on the superposition of the
Poisson distribution weighted by the gamma distribution,
so that the probability distribution is not an exact one.
By using the variational method, a tremendous reduction of the dimensionality has been achieved,
but there is no free lunch here;
the price paid is that we cannot obtain an exact probability distribution.
However, unlike the system size expansion method,
the variational method does not neglect the discrete characteristics of the chemical reaction systems,
and hence there is no need to take care of the zero-boundary effect as a special boundary condition.

\section{Example II: bivariate case}

In the next example, we treat a little complicated reaction system with two chemical substances.
In this case, the Hartree approximation is not valid,
and we will show that the method beyond the Hartree approximation
is necessary for small reaction systems with large fluctuation.

\subsection{Model}

We consider a reaction system constructed as
\begin{align}
\begin{array}{l}
\varnothing \xrightarrow{c_1} X,  \\
X+Y \xrightarrow{c_2} 2Y,  \\
Y \xrightarrow{c_3} \varnothing,  \\
\varnothing \xrightarrow{c_4} Y.  \\
\end{array}
\end{align}
The corresponding master equation is 
\begin{align}
&\frac{d}{dt} P_{n_x, n_y} = \notag \\
&\quad c_1 \left[ P_{n_x-1,n_y} - P_{n_x,n_y} \right] \notag \\
&\quad + c_2 \left[ (n_x+1) (n_y-1) P_{n_x+1,n_y-1} - n_x n_y P_{n_x,n_y} \right] \notag \\
&\quad + c_3 \left[ (n_y+1) P_{n_x,n_y+1} - n_y P_{n_x,n_y} \right] \notag \\
&\quad + c_4 \left[ P_{n_x,n_y-1} - P_{n_x,n_y} \right].
\end{align}
It is possible to make the corresponding deterministic rate equations,
and the averages of $n_x$ and $n_y$ obtained by the deterministic rate equations
are
$\langle n_x \rangle_\textrm{rate} = c_1 c_3 / (c_1 c_2 + c_2 c_4)$
and
$\langle n_y \rangle_\textrm{rate} = ( c_1 + c_4 ) / c_3$,
respectively.

The time evolution operator in the second quantized description is
\begin{align}
\Omega =& c_1 (a_x^\dagger - 1) + c_2 (a_x a_y^\dagger a_y^\dagger a_y - a_x^\dagger a_x a_y^\dagger a_y) \notag \\
&+ c_3 (a_y - a_y^\dagger a_y) + c_4 (a_y^\dagger - 1).
\end{align}

In the following subsections, we explain the Hartree approximation with two lognormal distributions
and a \textit{multivariate} superposition ansatz with a bivariate lognormal distribution.

\subsubsection{Hartree approximation}

In the Hartree approximation, 
we set the variational function by using two lognormal distributions as follows:
\begin{align}
| \Psi \rangle =& \int_0^\infty dx \int_0^\infty dy f(x; \mu_x, \sigma_x) f(y; \mu_y, \sigma_y) \notag \\
& \times \exp[ x (a_x^\dagger -1) ] \exp[ y (a_y^\dagger -1) ] | 0_x, 0_y \rangle,  \\
\langle \Phi | =& \langle 0_x, 0_y | \exp\left\{ a_x + \lambda_x^{(1)} a_x + \lambda_x^{(2)} (a_x)^2 \right. \notag \\
& \left. \qquad \qquad \qquad \quad + \lambda_y^{(1)} a_y + \lambda_y^{(2)} (a_y)^2\right\},
\end{align}
where $f(x;\mu_x,\sigma_x)$ and $f(y;\mu_y,\sigma_y)$ are lognormal distributions:
\begin{align}
f(x; \mu_x,\sigma_x) = \frac{1}{x \sigma_x \sqrt{2\pi}} \exp
\left[
- \frac{(\ln x - \mu_x)^2}{2 \sigma_x^2}
\right],
\end{align}
and $f(y;\mu_y,\sigma_y)$ is defined as the similar manner.
The first and second moments of the lognormal distribution are given by
\begin{align}
\langle x \rangle &= \exp\left[ \mu_x + \sigma_x^2 / 2\right], \\
\langle x^2 \rangle &= \exp\left[ 2 \mu_x + 2 \sigma_x^2\right].
\end{align}
In summary, the variational parameters are as follows:
\begin{align}
\begin{array}{l}
\alpha^\mathrm{R} = \{ \mu_x, \sigma_x, \mu_y, \sigma_y \}, \\
\alpha^\mathrm{L} = \{ \lambda_x^{(1)}, \lambda_x^{(2)}, \lambda_y^{(1)}, \lambda_y^{(2)} \}, \\
\end{array}
\end{align}
and then we obtain the following four coupled differential equations from Eq.~\eqref{eq_variational_equation}:
\begin{align}
&\frac{d \mu_x}{dt} \frac{\partial}{\partial \mu_x} \langle x \rangle
+\frac{d \sigma_x}{dt} \frac{\partial}{\partial \sigma_x} \langle x \rangle
= c_1 - c_2 \langle x y \rangle, \label{eq_bi_1} \\
&\frac{d \mu_y}{dt} \frac{\partial}{\partial \mu_y} \langle y \rangle
+\frac{d \sigma_y}{dt} \frac{\partial}{\partial \sigma_y} \langle y \rangle
= c_2 \langle xy \rangle - c_3 \langle y \rangle + c_4, \label{eq_bi_2}\\
&\frac{d \mu_x}{dt} \frac{\partial}{\partial \mu_x} \langle x^2 \rangle
+\frac{d \sigma_x}{dt} \frac{\partial}{\partial \sigma_x} \langle x^2 \rangle
= 2 c_1 \langle x \rangle - 2 c_2 \langle x^2 y \rangle, \label{eq_bi_3}\\
&\frac{d \mu_y}{dt} \frac{\partial}{\partial \mu_y} \langle y^2 \rangle
+\frac{d \sigma_y}{dt} \frac{\partial}{\partial \sigma_y} \langle y^2 \rangle \notag \\
&\quad = 2 c_2 \langle xy^2 \rangle + 2c_2 \langle xy \rangle  - 2 c_3 \langle y^2 \rangle + c_4 \langle y \rangle. 
\label{eq_bi_4}
\end{align}
Since we here use the Hartree approximation,
$x$ and $y$ do not correlate.
Hence, we set
$\langle x y \rangle = \langle x \rangle \langle y \rangle$,
$\langle x^2 y \rangle = \langle x^2 \rangle \langle y \rangle$,
and so on.

\subsubsection{Bivariate lognormal distribution}

In the reaction system with two chemical substances, 
it is expected that the correlation between $n_x$ and $n_y$ plays an important role,
so that we must go beyond the Hartree approximation.
While the Poisson ansatz needs the Hartree approximation,
the superposition ansatz enables us to treat the correlation between $n_x$ and $n_y$.
In order to treat such correlations,
we here use the bivariate lognormal distribution:
\begin{widetext}
\begin{align}
f(x,y ; \mu_x, \mu_y, \sigma_x, \sigma_y, \rho) 
=& \frac{1}{2 \pi x y \sigma_x \sigma_y \sqrt{1-\rho^2} } \notag \\
& \times \exp\left[ 
-\frac{1}{2} \left( \frac{1}{1-\rho^2}\right)
\left\{ 
\left(\frac{\ln x - \mu_x}{\sigma_x}\right)^2 
- 2 \rho \left( \frac{\ln x - \mu_x}{\sigma_x}\right) \left( \frac{\ln y - \mu_y}{\sigma_y}\right)
+ \left(\frac{\ln y - \mu_y}{\sigma_y}\right)^2 
\right\}
\right].
\end{align}
\end{widetext}
The ket ansatz and bra ansatz are set as
\begin{align}
| \Psi \rangle =& \int_0^\infty dx \int_0^\infty dy f(x,y; \mu_x, \sigma_x, \mu_y, \sigma_y, \rho) \notag \\
& \times \exp[ x (a_x^\dagger -1) ] \exp[ y (a_y^\dagger -1) ] | 0_x, 0_y \rangle,
\end{align}
and
\begin{align}
\langle \Phi | =& \langle 0_x, 0_y | \exp\left\{ a_x + \lambda_x^{(1)} a_x + \lambda_x^{(2)} (a_x)^2 \right. \notag \\
& \left. \qquad \qquad \qquad + \lambda_y^{(1)} a_y + \lambda_y^{(2)} (a_y)^2 + \lambda_{xy} a_x a_y\right\}.\,
\end{align}
respectively.
By using the bivariate lognormal distribution,
we can easily obtain the following quantities which are necessary to evaluate the time evolution equations:
\begin{align}
\langle x y \rangle &= \exp\left[ \rho \sigma_x \sigma_y 
+ \mu_x + \sigma_x^2 / 2 + \mu_y + \sigma_y^2 / 2\right], \\
\langle x^2 y \rangle &= \exp\left[ 2 \rho \sigma_x \sigma_y 
+ 2 \mu_x + 2 \sigma_x^2 + \mu_y + \sigma_y^2 /2 \right], \\
\langle x y^2 \rangle &= \exp\left[ 2 \rho \sigma_x \sigma_y 
+ \mu_x + \sigma_x^2 /2  + 2 \mu_y + 2 \sigma_y^2  \right]. 
\end{align}
The variational parameters are
\begin{align}
\begin{array}{l}
\alpha^\mathrm{R} = \{ \mu_x, \sigma_x, \mu_y, \sigma_y, \rho \}, \\
\alpha^\mathrm{L} = \{ \lambda_x^{(1)}, \lambda_x^{(2)}, \lambda_y^{(1)}, \lambda_y^{(2)}, \lambda_{xy} \}, \\
\end{array}
\end{align}
and then we obtain totally five coupled differential equations;
four equations are equivalent to Eqs.~\eqref{eq_bi_1}-\eqref{eq_bi_4},
and we obtain an additional equation:
\begin{align}
&\frac{d \mu_x}{dt} \frac{\partial}{\partial \mu_x} \langle xy \rangle
+\frac{d \sigma_x}{dt} \frac{\partial}{\partial \sigma_x} \langle xy \rangle
+\frac{d \mu_y}{dt} \frac{\partial}{\partial \mu_y} \langle xy \rangle \notag \\
&\quad  +\frac{d \sigma_y}{dt} \frac{\partial}{\partial \sigma_y} \langle xy \rangle
+\frac{d \rho}{dt} \frac{\partial}{\partial \rho} \langle xy \rangle \notag \\
&= c_1 \langle y \rangle + c_2 \langle x^2 y \rangle - c_2 \langle xy \rangle - c_2 \langle x y^2 \rangle
- c_3 \langle xy \rangle + c_4 \langle x \rangle.
\end{align}

\subsection{Numerical result}

\begin{figure}
\begin{center}
  \includegraphics[width=70mm,keepaspectratio,clip]{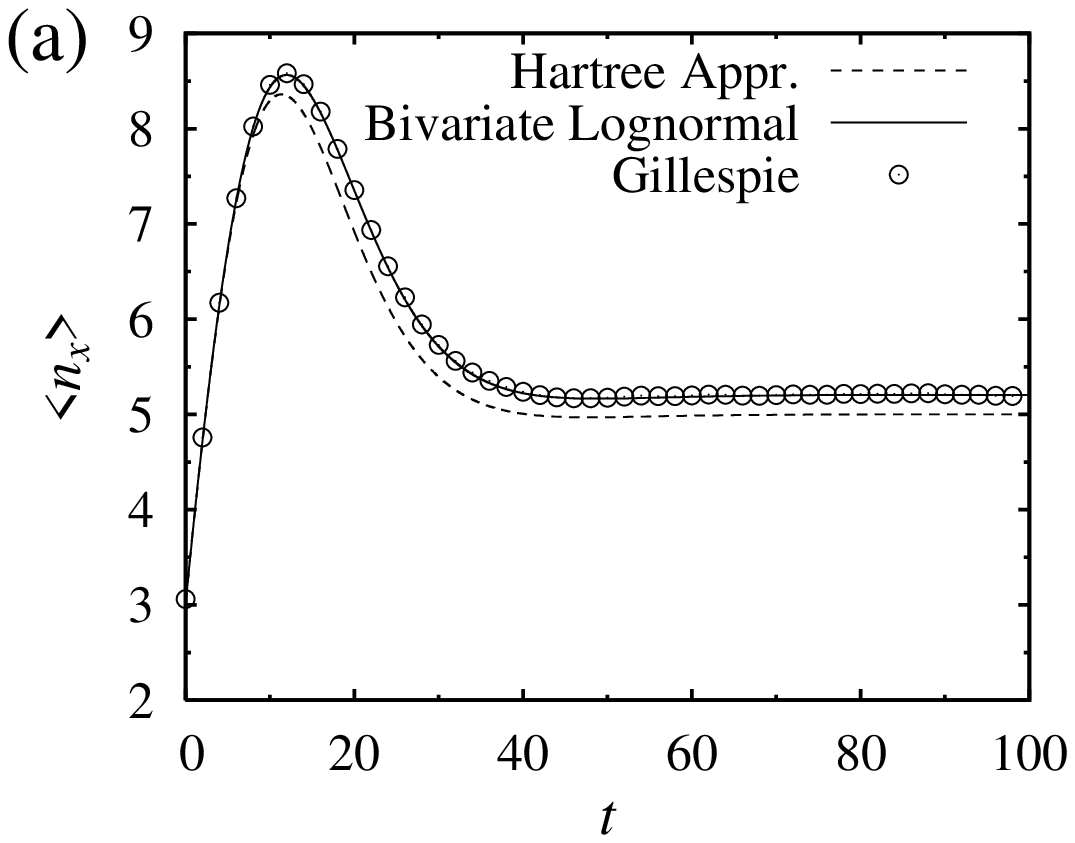} \\
  \includegraphics[width=70mm,keepaspectratio,clip]{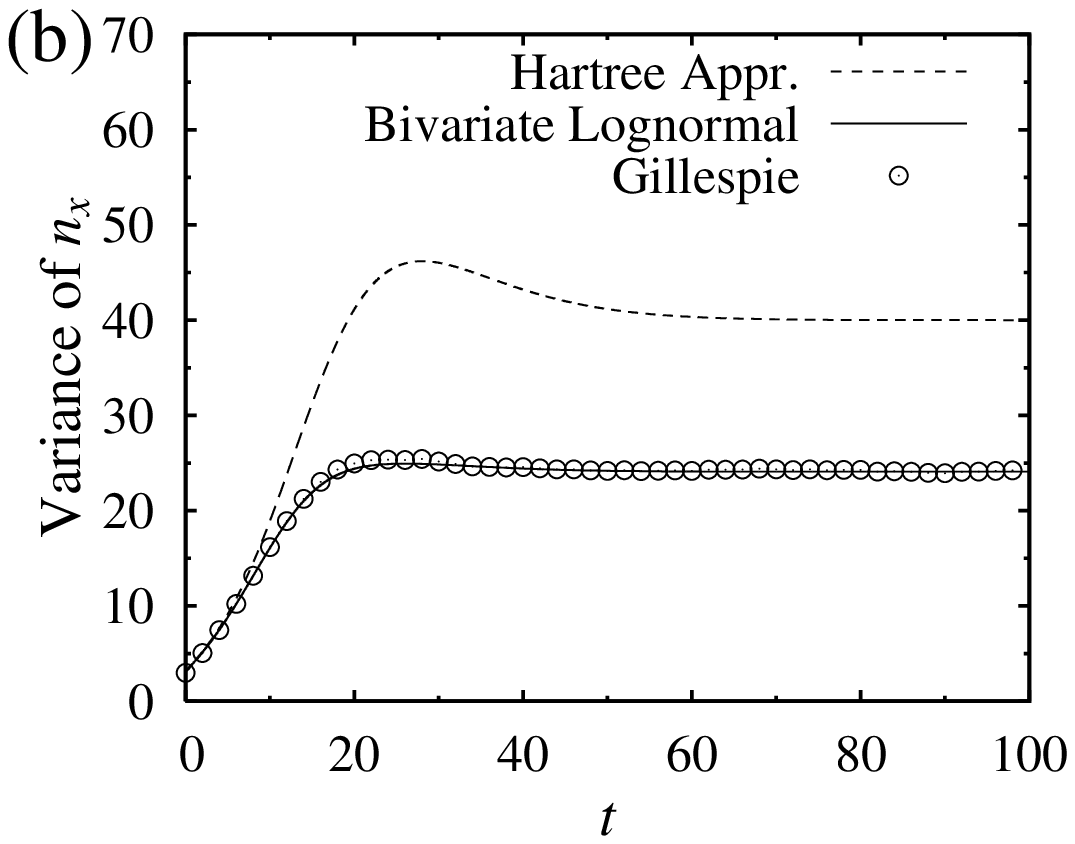} \\
  \includegraphics[width=70mm,keepaspectratio,clip]{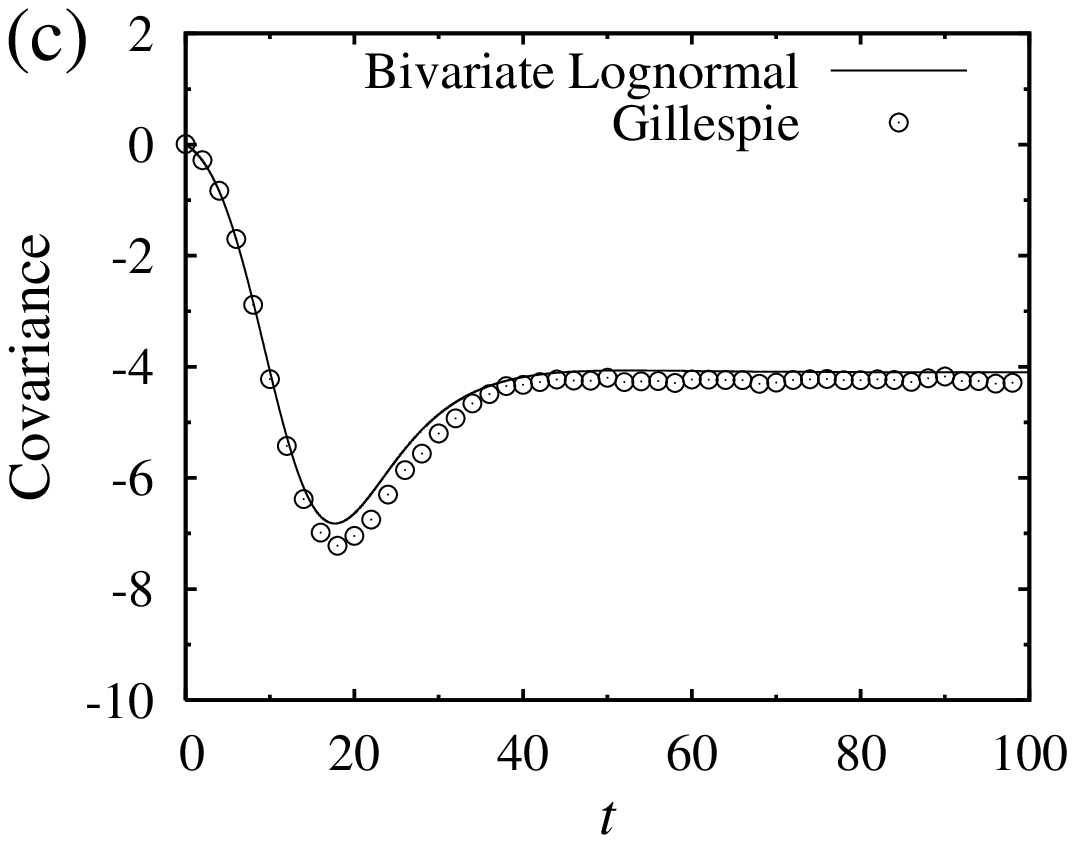}
\caption{
(a) Time evolution of the average of $n_x$.
The results obtained by the Hartree approximation slightly deviates from the results by the Gillespie algorithm.
(b) Time evolution of the variance of $n_x$.
(c) Time evolution of the covariance.
Here, we set 
$c_1 = c_4 = 1$, $c_2 = 0.01$, $c_3 = 0.1$.
}
\label{fig_bi_result_1}
\end{center}
\end{figure}

We performed numerical experiments in order to clarify the advantage and application limitation 
of the variational method.
Here, we set 
$c_1 = c_4 = 1$, $c_2 = 0.01$, $c_3 = 0.1$ and 
initial parameters are set to $\mu_x = \mu_y = \log 3$, $\sigma_x = \sigma_y = 0.1$,
and for the bivariate lognormal distribution, $\rho = 0$.
As in the case of Sec.~III.A, the numerical results obtained by the Gillespie algorithm are averaged over $10^5$ runs.

Figure~\ref{fig_bi_result_1} shows the time evolution of the mean of $n_x$.
The mean of $n_x$ is obtained by 
\begin{align}
\langle n_x \rangle &\equiv \sum_{n_x = 0}^\infty \sum_{n_y = 0}^\infty n_x 
\int_0^\infty dx \int_0^\infty dy f(x,y;\mu_x, \mu_y, \sigma_x, \sigma_y, \rho) \notag \\
&\times \frac{e^{-x} x^{n_x}}{n_x!} \frac{e^{-y} y^{n_y}}{n_y!} \notag \\
&= \sum_{n_x = 0}^\infty  n_x \int_0^\infty dx f(x,y;\mu_x, \mu_y, \sigma_x, \sigma_y, \rho)
\frac{e^{-x} x^{n_x}}{n_x!} \notag \\
&= \langle x \rangle,
\end{align}
for the multivariate superposition ansatz.
For the Hartree approximation scheme,
we have the same relationship; $\langle n_x \rangle = \langle x \rangle$.
From Fig.~\ref{fig_bi_result_1}(a),
one can see that the result obtained by the Hartree approximation 
slightly deviates from the results of the Gillespie algorithm.
From the deterministic rate equations,
we predict the mean as $\langle n_x \rangle_\textrm{rate} =  5$.
Hence, the results by the Gillespie algorithm and the bivariate lognormal distribution
suggest that the correlation between $n_x$ and $n_y$ varies the mean value of $n_x$
from that of the deterministic rate equation approach.
Figure~\ref{fig_bi_result_1}(b) shows the variance of $n_x$.
$\textrm{Var}[n_x]$ is calculated from the same procedure as Eq.~\eqref{eq_variance}.
As in the case of the mean,
the results obtained by the bivariate lognormal distribution are quantitatively in good agreement
with the Monte Carlo results.
In addition, the covariance between $n_x$ and $n_y$ cannot be calculated
by the Hartree approximation and, of course, the Poisson ansatz.
The superposition ansatz enables us to calculate the covariance
by $\textrm{Cov}[n_x,n_y] = \langle x y \rangle - \langle x \rangle \langle y \rangle$.
The result by the bivariate lognormal distribution is quantitatively in good agreement with that of
the Gillespie algorithm in the present case, as shown in Fig.~\ref{fig_bi_result_1}(c).

\begin{figure}
\begin{center}
  \includegraphics[width=70mm,keepaspectratio,clip]{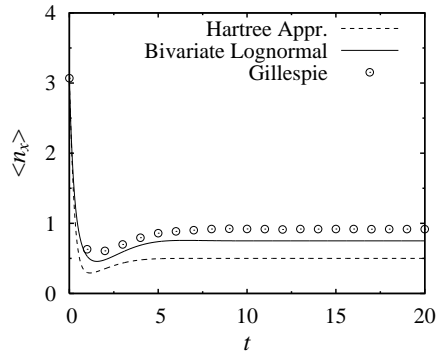}
\caption{
Time evolution of the average $n_x$.
We here set $c_1 = c_2 = c_3 = 1$.
}
\label{fig_result_bi_2}
\end{center}
\end{figure}

We showed that the variational method beyond the Hartree approximation is useful
for calculating physical quantities, especially for small systems in which
correlations and fluctuations play important roles.
By using the variational method, there is no need to take care of the zero-boundary, as explained in Sec.~III.
Of course, it is needed to restrict the variational function into some specific form,
and then it could be difficult to treat the correlations and fluctuations correctly in some cases.
Figure~\ref{fig_result_bi_2} shows the time evolution of the average $n_x$
in the case with $c_1 = c_2 = c_3 = 1$.
The average $n_x$ derived by the Hartree approximation
is largely different from the results by the Gillespie algorithm.
In addition, the results by the bivariate lognormal distribution also deviates form that of the Gillespie algorithm,
although the deviation is small compared with the case of the Hartree approximation.
This means that in such small $n_x$ region, the higher correlation should be taken into account.
We expect that the usage of more complicated variational function enables us to improve the deviation.

\section{Concluding remarks}

In summary, 
we clarified that it is possible to apply the superposition ansatz
to multivariate cases without the usage of the Hartree approximation.
The superposition ansatz enables us to treat various correlations,
and it gives better numerical results than the Poisson ansatz.
Such correlations can be included by adding only a few differential equations (in our case, only one equation is added),
and then the computational cost increases very little.
In addition, we showed that the survival probability is easily calculated via the Laplace transformation.

In the present paper, we considered a reaction system with two chemical species,
but in principle, the superposition ansatz is applicable to any multivariate case.
In addition, we believe that 
the second-quantization scheme has a possibility to create more powerful analytical method,
with the aid of various techniques developed in the quantum physics.

Although we have checked that the multivariate lognormal distribution is easy to treat
in the variational method,
it will sometimes be necessary to use more complicated probability distributions as the variational function.
As future works, such improvement of the variational function is needed.


\end{document}